\documentclass[12pt]{article}
\usepackage[polish,english]{babel}
\usepackage[latin1]{inputenc}
\usepackage{times}
\usepackage[T1]{fontenc}
\usepackage{epsfig}

\begin{document}

\title{Compact oscillons in the signum-Gordon model }

\author{H. Arod\'z, $\;\;$ P. Klimas $\;$ and $\;$ T. Tyranowski \\$\;\;$ \\  Institute of Physics,
Jagiellonian University, \\ Reymonta 4, 30-059 Cracow, Poland}

\date{$\;$}

\maketitle

\begin{abstract}
We present  explicit solutions of the signum-Gordon scalar field equation which have finite energy and are
periodic in time.  Such oscillons have strictly finite size. They do not emit radiation.
\end{abstract}

\vspace*{2cm} \noindent PACS: 05.45.-a, 03.50.Kk, 11.10.Lm \\
\noindent Preprint TPJU - 5/2007

\pagebreak

\section{ Introduction}

Dynamics of self-interacting scalar fields is still a subject of very interesting and important research.
Because such fields are ubiquitous in physics, from condensed matter to cosmology, the results can have a wide
impact on our understanding of the material world. Nonlinearity of pertinent field equations is the reason for
the existence of very nontrivial and often unexpected phenomena. In particular, much attention has been paid to
so called oscillons, extremely long-lived scalar field configurations which in many respects behave like
solitons, but which finally decay into radiation, see e.g. recent papers \cite{1}. The nonlinearity can have
various forms. In the current literature dominate models with polynomial Lagrangians. However, in a recent paper
\cite{2} we have pointed out a rather interesting class of scalar field models with non polynomial field
potentials which are V-shaped at their minima and which have another kind of nonlinearity. The simplest example
of such models is the signum-Gordon model.

The signum-Gordon model (s-G) has the following  Lagrangian \cite{3}
\begin{equation} L = \frac{1}{2} \frac{\partial \varphi}{\partial x^{\mu}} \frac{ \partial \varphi}{\partial x_{\mu}}
 - g\:|\varphi|, \end{equation} where $\varphi$ is a real (1+1)-dimensional scalar field, $\mu =0, 1,$ and
$x^0, x^1$ are dimensionless variables obtained by the appropriate choice of units for the physical time and
position coordinates.  $g>0$ is a dimensionless coupling constant.  The pertinent field potential has the form
\[
U(\varphi) = g \:|\varphi|.\] Its left and right derivatives  at $\varphi =0,$ where it has the absolute
minimum, do not vanish -- they are equal to $\mp g$. Hence, in case of the s-G model the second derivative
$U''(0)$ is (in a sense) infinite. In majority of field theoretic models the field potential $U(\varphi)$ is
smooth at its absolute minimum at $\varphi = \varphi_0$, and $ \lambda_0 = (U''(\varphi_0))^{-1/2}$ is
recognized as the fundamental length scale in the model. In the case of massless models $\lambda_0$ is infinite,
while finite $\lambda_0 >0$ characterizes massive models with the finite mass parameter $m_0^2 = 1/\lambda_0^2$.
In this classification, the s-G model corresponds to $\lambda_0=0$ and infinite $m_0^2$ -- for this reason it
can be called a supermassive model.

Several rather interesting features of the models with V-shaped field potentials have been pointed out in our
previous papers: the existence of static, compact solitons (topological compactons) \cite{4} ,  the presence of
a scaling symmetry \cite{2}, and the variety of exact self-similar solutions \cite{3, 5}. In these papers we
have also presented a sound physical motivation for considering such models. It includes the description of an
infinite chain of harmonically coupled classical pendulums, and the dynamics of an elastic string (a vortex)
pinned with a constant force to a line. Because of such a down-to-earth physics behind the s-G model, it is not
surprising that the model is perfectly well-behaved from the physical viewpoint. In particular, the conserved
energy is bounded from below, and the field equation is of the standard hyperbolic type with the usual causality
features. Nevertheless, the dynamics of the scalar field in this model is strongly influenced by the fact that
the field potential is not smooth at the minimum. Moreover, the model cannot be linearized even if the amplitude
of the scalar field is arbitrarily small, contrary to the models with smooth field potentials. Perhaps it is one
of the reasons for which the models with V-shaped field potentials were not discussed in literature.

In the present paper we show that the s-G model possesses solutions with finite energy which are periodic in
time -- the oscillons. These oscillons are rather interesting for the following reasons. First, they are given
explicitly as exact solutions of the s-G field equation. Moreover,  they do not loose their energy because they
do not emit any  radiation. They have a compact support in the space, i.e., a strictly finite size. In these
respects they differ from the oscillons of the $\varphi^4$ model which are known only as approximate numerical
solutions and seem to contain a radiative component. Because of the scaling invariance of the s-G equation in
fact there is a continuous family of oscillons with arbitrary positive energies. One can also trivially combine
such oscillons to form multi-oscillon solutions.

\section{The signum-Gordon equation }

The signum-Gordon equation is the Euler-Lagrange equation
corresponding to Lagrangian (1). It can be written in the
following form
\begin{equation}
\frac{\partial^2 \varphi(x, t)}{\partial t^2} = \frac{\partial^2 \varphi(x, t)}{\partial x^2}  -
\mbox{sign}(\varphi(x, t)),
\end{equation}
where $t= \sqrt{g} x^0, \: x = \sqrt{g} x^1.$ The $\mbox{sign}$ function has the values $\pm 1$ when $ \varphi
\neq 0$ and 0 for $ \varphi = 0.$ Because of the $- \mbox{sign}(\varphi)$ term Eq. (1) is nonlinear in a rather
special way. Mathematical aspects of this equation are discussed  in \cite{3}. Generally speaking, physically
relevant are solutions which are continuous functions of $x$ and $t$, but only piecewise smooth. They belong to
the class of so called weak solutions of partial differential equations.  The reader not familiar with the
mathematics of such solutions should consult the literature, e.g., \cite{6, 7}.

The novel feature of Eq. (2) is that the nonlinear term  $-\mbox{sign}(\varphi)$ remains finite for arbitrarily
small $|\varphi|$ provided that $\varphi \neq 0.$ This fact has profound influence on the dynamics of weak
fields. In other models nonlinear terms vanish when fields become arbitrarily small, cf. the $-\varphi^3$
nonlinear term in the case of the nonlinear Klein-Gordon equation. Consider, for example, the Gaussian wave
packet $\varphi(x, t=0) = \exp(-x^2)$ at the initial time $t=0$. Let us assume for simplicity that
$\partial_t\varphi(x, t=0)=0.$ The field acceleration  $\partial^2_t \varphi(x, t=0)$ is obtained from the field
equations. In the case of the $-\varphi^3$ nonlinearity the dominating  at large $|x|$ contribution to the
acceleration comes from the $ \partial^2_x \varphi(x, t=0)$ term and it is positive, hence the Gaussian wave
packet will spread out. In the case of the s-G equation the $-\mbox{sign}(\varphi)$ term dominates at large
$|x|$ and it is negative, hence the wave packet will shrink for certain time. Heuristically, both `forces'
$-\varphi^3$ and $-\mbox{sign}(\varphi)$ push the field toward the equilibrium value $\varphi =0$, but the
$-\mbox{sign}(\varphi)$ force is much more effective in this respect.

The signum term in Eq. (2) remains constant until $ \varphi$ becomes equal to zero. Therefore,  on each interval
on the $x$ axis such that $\varphi$ has a constant sign on it, one can use the well-known formula for the
general solution of the one dimensional wave equation, suitably modified to incorporate the constant $+1$ or
$-1$ term in the equation:
\begin{equation}
\varphi(x,t) = h(x-t) + g(x+t) + c_1 t^2,
\end{equation}
where $ 2 c_1  = \pm 1 (= - sign(\varphi)). $ The functions $ h, g$ and the constant $c_1 $ are determined from
the initial and  boundary conditions.

\section{The oscillon }

Motivated by the heuristic considerations about the wave packets
we shall try to construct a solution of the s-G equation (2) which
does not spread out at least during a certain time interval. We
start by specifying simple initial data at $t=0$:
\begin{equation}
\varphi(x, 0)=0, \;\;\;\partial_t\varphi(x, 0)=\left\{
\begin{array}{lcl} 0 & \;\;\; \mbox{if} \;\;\; & x \leq 0, \\ v(x)& \;\;\; \mbox{if} \;\;\;
&  0<x<1, \\   0 & \;\;\; \mbox{if} \;\;\; & x \geq 1,
\end{array} \right.
\end{equation}
where the initial field velocity $v(x)$ is assumed to be negative. Its exact form will be determined later.
Moreover, we impose the following boundary conditions
\begin{equation}
\varphi(x=0, t)=0, \;\;\; \varphi(x=1, t)=0.
\end{equation}
The hope is that the field will remain localized in the interval $0 \leq x \leq 1$ for arbitrary times $t >0.$

Because $ v(x)$ is negative, for small enough $t >0$ the field $\varphi$ will be negative or equal to zero.
Therefore, in the region $ 0 < x <1$ we use formula (3) in which  $ c_1 = 1/2.$  The functions $h$ and $g$ are
determined from conditions (4), (5). For a certain reason which will become clear later we denote the solution
given below by $ \varphi^-$. Simple calculations give the following results
\begin{equation}
\varphi^-(x,t) = \left\{ \begin{array}{lcl} \varphi_1(x,t) & \;\;\; \mbox{if} \;\;\; &0 \leq x \leq t, \\
\varphi_2(x,t) & \;\;\; \mbox{if} \;\;\; & t \leq x \leq 1-t, \\
\varphi_3(x,t)  & \;\;\; \mbox{if} \;\;\; & 1-t \leq x \leq 1,
\end{array} \right.
\end{equation}
where
\begin{eqnarray}
\varphi_1(x,t)& = & - \frac{x^2}{2}  + tx + \frac{1}{2}
\int^{t+x}_{t-x} \!\!\! ds \: v(s), \\
\varphi_2(x,t)&= & \frac{t^2}{2} + \frac{1}{2} \int^{t+x}_{x -t}\!\!\!
ds \: v(s),\\
\varphi_3(x,t)& =  & - \frac{x^2}{2} - \frac{1}{2} - x t + x +t + \frac{1}{2} \int^{2 - t -x}_{x-t}\!\!\! ds \:
v(s).
\end{eqnarray}

The reason for splitting the unit interval $0 \leq x \leq 1$ into the three parts is that the boundary
conditions (5) modify the field at the points which are causally connected with the points $x=0$ and $ x=1$:
such points form the two subintervals $0 \leq x \leq t, \;\;  1-t \leq x \leq 1.$ These subintervals meet at the
time $t = 1/2$. Hence, the solution given above is valid for $t$ in the interval $[0, 1/2]$, and for $x$ in the
interval  $ [0, 1].$

 Note that at the initial time $t=0$ only $\varphi_2$ is present because the spatial
supports of $ \varphi_1, \: \varphi_3$ are shrank to the points $x=0, \: x=1$, respectively. The functions
$\varphi_1, \: \varphi_2$ match each other at the point $x=t$, similarly as $\varphi_2, \: \varphi_3$ at $x=
1-t$, so that $\varphi^-$ as well as $\partial_t\varphi^-, \; \partial_x \varphi^-$ are continuous functions of
$x, t$.

In the next step we extend the solution  $\varphi^-$ to the whole $x$ axis. Of course we just would like  to put
$\varphi(x,t) =0 $ for $x <0$ and for $x >1.$ This is possible provided that
\begin{equation}
\partial_x\varphi_1(x=0,t) =0, \;\;\;  \partial_x\varphi_3(x=1,t) =0,
\end{equation} because otherwise $\partial_x\varphi$ would have
discontinuities at the points $x=0,\: x=1.$  Such discontinuities, if present, would have to move with `the
velocity of light'  $\pm1$ because the  s-G equation belongs to the class of hyperbolic partial differential
equations. Conditions (10) give the following equations
\[
v(t) = -t, \;\;\; v(1-t) = -t,
\]
where $ 0 \leq t \leq 1/2$. It follows that the initial field
velocity $v(x)$  has the following form
\begin{equation}
v(x) = \left\{  \begin{array}{lcl} -x & \;\;\; \mbox{if} \;\;\; &
0 \leq x \leq \frac{1}{2}, \\ x-1 & \;\;\; \mbox{if} \;\;\; &
\frac{1}{2} \leq x \leq 1, \end{array} \right.
\end{equation}
see also Fig. 1.
\begin{center}
\begin{figure}[tph!]
\hspace*{1cm}
\includegraphics[height=6.0cm, width=10cm]{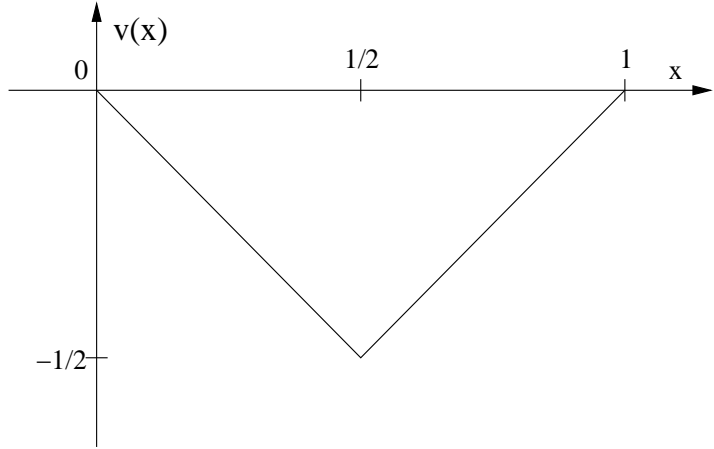}
\caption{The initial profile of $\partial_t \varphi$ for the oscillon solution. }
\end{figure}
\end{center}

Now we can compute the integrals present in formulas (7-9). It is
convenient to split the time interval [0, 1/2] into two parts. It
turns out that when $t$ has values from the interval $ [0, 1/4]$
then
\begin{eqnarray}
\varphi_1(x,t)& = & - \frac{x^2}{2}, \\
\varphi_2(x,t)&= &\left\{ \begin{array}{lcl}  \frac{t^2}{2} - x t
& \;\;\;\mbox{if} \;\;\; &  t \leq x \leq \frac{1}{2} -t,  \\
\frac{x^2}{2} + t^2  - \frac{x}{2} - \frac{t}{2}+ \frac{1}{8}
& \;\;\; \mbox{if} \;\;\; &  \frac{1}{2} -t \leq x \leq \frac{1}{2} + t, \\
\frac{t^2}{2} + t (x-1) & \;\;\; \mbox{if} \;\;\; & \frac{1}{2} +
t \leq x \leq 1-t, \end{array} \right. \\
\varphi_3(x,t)&= &- \frac{(1-x)^2}{2},
\end{eqnarray}
and for $t$ from the interval $ [1/4, 1/2]$
\begin{eqnarray}
\varphi_1(x,t)& = &\left\{ \begin{array}{lcl} - \frac{x^2}{2}
& \;\;\;\mbox{if} \;\;\; &  0 \leq x \leq \frac{1}{2} -t,  \\
\frac{t^2}{2} + t x - \frac{x}{2} - \frac{t}{2}  + \frac{1}{8}& \;\;\; \mbox{if} \;\;\; &  \frac{1}{2} -t \leq x
\leq  t,
\end{array} \right. \\
\varphi_2(x,t) & = & t^2+ \frac{x^2}{2}   -
\frac{x}{2} - \frac{t}{2}+ \frac{1}{8},\\
\varphi_3(x,t)& = &\left\{ \begin{array}{lcl} \frac{t^2}{2} - t x
 + \frac{x}{2} + \frac{t}{2} -\frac{3}{8}
& \;\;\;\mbox{if} \;\;\; &  1-t \leq x \leq \frac{1}{2} + t,  \\
  - \frac{(1-x)^2}{2}& \;\;\; \mbox{if} \;\;\; &  \frac{1}{2} + t \leq
x \leq  1.
\end{array} \right.
\end{eqnarray}
Note that  $\partial_t\varphi(x,t) =0$ at the time $t=1/4$.

To summarize, we have found a solution of the s-G equation such that $\varphi =0$ outside the interval $[0,1]$
on the $x$-axis, and inside this interval it has the form given by formulas (6),  (12-17), provided that $0 \leq
t \leq 1/2.$ Snapshots of the field configurations at the times $t=1/8$ and $t=1/4$ are presented in Figs. 2 and
3.
\begin{center}
\begin{figure}[tph!]
\hspace*{1cm}
\includegraphics[height=6.0cm, width=10cm]{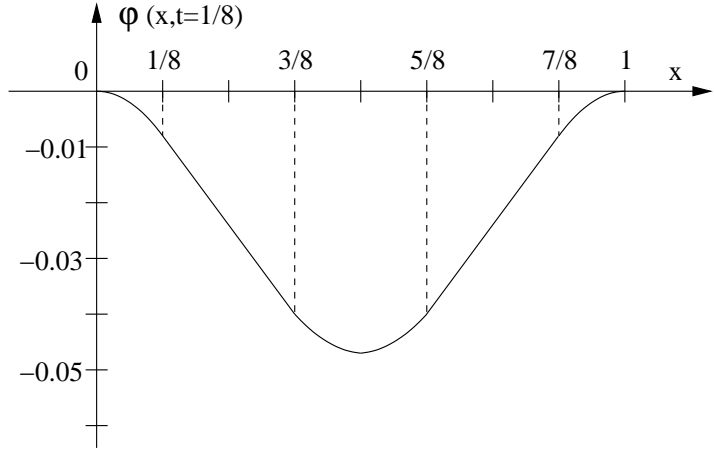}
\caption{Picture of the oscillon solution at the time $t=1/8$. For $x$ from the intervals [1/8, 3/8] and [5/8,
7/8] the function $\varphi(x, 1/8) $ is represented by rectilinear segments which  at the time $t=1/4$ shrink to
the points 1/4 and 3/4, respectively, see the first and third lines of formula (13).}
\end{figure}
\end{center}
\begin{center}
\begin{figure}[tph!]
\hspace*{1cm}
\includegraphics[height=6.0cm, width=10cm]{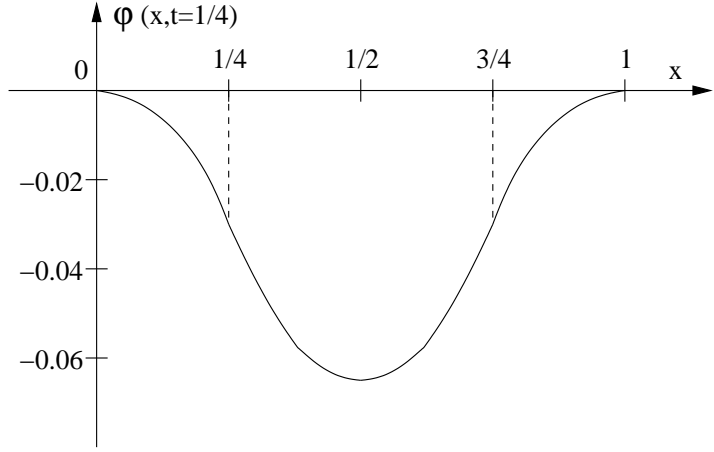}
\caption{Picture of the oscillon solution at the time $t=1/4$. This configuration is `the turning point' because
precisely at this time $\partial_t\varphi$  changes its sign from negative to positive. }
\end{figure}
\end{center}
For times $t$ from the interval $(1/4, 1/2)$ the solution has the same shape as presented in Fig. 2, but now the
two rectilinear segments expand and move until they cover the intervals $[0, 1/2], \; [1/2,1]$ on the $x$-axis.

In order to find $\varphi$ for times larger than 1/2 we compute $\varphi^-$ and $\partial_t\varphi^-$ at $t=1/2$
and we take these functions of $x$ as the new initial data for the s-G equation. The calculation is trivial, but
one should  remember about watching the domains of the involved functions because they are given by inequalities
which also depend on time. The result is quite surprising and crucial for further progress:
\begin{equation}
\varphi^-(x, t=1/2) =0, \;\;\; \partial_t\varphi^-(x, t=1/2) = -
v(x),
\end{equation}
where $v(x) $ is given by formula (11). These formulas imply that
$\varphi(x,t) $ for $t>1/2$ can be obtained with the help of
symmetries of the s-G equation: the time translation $t
\rightarrow t-1/2$ and the change of sign of the field $\varphi
\rightarrow - \varphi$. The solution obtained in this way is
denoted by $\varphi^+$. Thus,
\begin{equation}
\varphi^+(x, t) = - \varphi^-(x, t-\frac{1}{2}).
\end{equation}
The solution $\varphi^+(x, t)$ holds for $ t$ in the interval $[1/2, 1]$. At the time $t=1$ the field $\varphi$
and its time derivative $\partial_t \varphi$ return to their initial values (4). Thus, the period of the
oscillon  is equal to its spatial size, i.e., to 1.

The structure of the oscillon solution during one period is schematically depicted in Fig. 4. The functions
$\varphi^{\pm}_{\alpha}$, where $\alpha = 1A, 1B, ...$, are just parts of formulas (12)-(17) -- the new notation
is introduced for the clarity of the figure. Let us list these functions for the convenience of the reader:
\[
\varphi^-_{1A} = - \frac{x^2}{2}, \;\;\; \varphi_{1B}^- =
\frac{t^2}{2} + tx - \frac{t}{2} - \frac{x}{2} + \frac{1}{8},
\] \[ \varphi_{2A}^- = \frac{t^2}{2} - tx, \;\;\; \varphi_{2B}^- =
\frac{t^2}{2} + t(x -1), \;\;\;\varphi_{2C}^- = t^2 +  \frac{x^2}{2}  - \frac{x}{2} -\frac{t}{2} + \frac{1}{8},
\]
\[\varphi_{3A}^- =
\frac{t^2}{2} - tx + \frac{t}{2} +\frac{x}{2} - \frac{3}{8}, \;\;\;\varphi_{3B}^- = - \frac{1}{2}(x-1)^2, \] and
the functions $ \varphi_{\alpha}^+(x,t) $ are obtained from formula (19). Note that the oscillon is spatially
symmetric:  the solution is invariant under the spatial reflection at the point $x=1/2$, that is with respect to
the substitution $x \rightarrow 1-x$. The point $x=1/2$ is the center of the oscillon.

\begin{center}
\begin{figure}[tph!]
\hspace*{1.5cm}
\includegraphics[height=7.0cm, width=10cm]{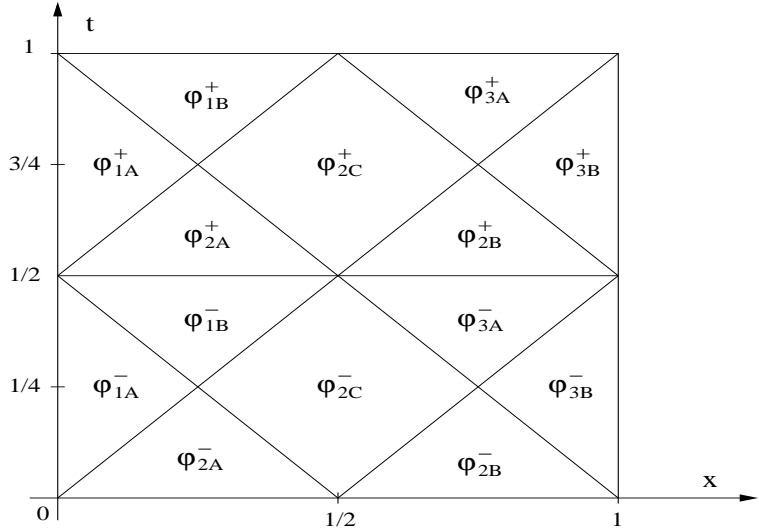}
\caption{Schematic picture of the oscillon solution. Because of the  periodicity in time,  the figure from the
unit square should be repeated for $t<0$ and for $t>1$. }
\end{figure}
\end{center}

The s-G equation is dilation invariant \cite{3}: if $\varphi(x,t)$ is a solution of it, than so is also
\[
\varphi_l(x,t) = l^2\varphi(\frac{x}{l}, \frac{t}{l}),
\]
where $l$ is an arbitrary real, positive constant. Applying this symmetry to the oscillon solution presented
above we generate a one-parameter family of oscillons of the length $l/\sqrt{g}$ with the period  equal to the
length (let us recall that we use the dimensionless variables $x^{\mu}$). The amplitude of the oscillations of
the scalar field  is equal to $l^2/16$. The total energy of the oscillons is given by the formula obtained from
Lagrangian (1)
\[
E = \frac{1}{2} \sqrt{g} \int^{\infty}_{-\infty}dx\; [(\partial_t\varphi)^2 + (\partial_x\varphi)^2]+ \sqrt{g}
\int^{\infty}_{-\infty}dx\; |\varphi| = \frac{\sqrt{g} }{24} l^3.
\]
The frequency of the oscillations of the field is $\sim l^{-1}$, hence the energy $E$ vanishes in the limit of
high frequencies.

The s-G equation is also invariant with respect to Poincar\'e transformations. In particular, boosts provide
oscillons moving with an arbitrary constant velocity $v$ such that $|v| <1$.

\section{Remarks}

\noindent 1. Note that the s-G oscillons have a strictly finite size, and that the field  approaches its vacuum
value $\varphi=0$ in the parabolic manner, see the functions $\varphi_{1A}^-, \;\varphi_{3B}^-$. In these
respects they are similar to the static topological compactons discussed in \cite{4} in another model with a
V-shaped potential. The compactness and the parabolic approach to the vacuum field seem to be generic features
of the models with V-shaped field potentials.

\noindent 2. The s-G equation admits explicit solutions which describe arbitrary chains of static, non
overlapping oscillons of arbitrary sizes. The multi-oscillon solutions are obtained trivially by adding
appropriately translated in space single oscillon solutions. Such oscillons do not interact with each other
because they have strictly finite sizes.

\noindent 3. We have attempted to investigate the stability of the s-G oscillon under small perturbations. Our
numerical simulations did not show any instability. On the analytic side, the problem  turns out to be quite
difficult. The main reason is that there is no  linear approximation  to the function $\mbox{sign}(\varphi)$
around $\varphi=0$. This fact seems to render the standard linear stability theory useless. Note however that
for $\varphi \neq 0$ and the perturbation $\varepsilon$ small enough
\[
\mbox{sign}(\varphi + \varepsilon)   = \mbox{sign}(\varphi).
\]
Therefore, in such cases the s-G equation implies the simple massless wave equation for the perturbation,
\[
\partial_{\mu}\partial^{\mu}\varepsilon =0.
\]
It follows that the perturbation does not grow: it  splits into a left- and right-movers which just travel with
the velocities $\pm1$ until they reach a point at which the function $\mbox{sign}(\varphi + \varepsilon)$
changes its sign. It remains to be investigated what happens at such points. Generally, one could expect
complicated dynamical processes including an emission of radiation  or of small oscillons.

\noindent 4. We expect that the s-G oscillons will likely have to be taken into account when considering various
dynamical processes with the s-G field.  The oscillons with arbitrarily small energies can probably be emitted
in such processes, thus forming a kind of 'infrared' cloud consisting of a number of small, compact oscillons.

It is clear that the dynamics of the s-G field turns out to be very exciting.

\section{Acknowledgement}
This work is supported in part by the project SPB nr. 189/6.PRUE/2007/7.

\end{document}